\documentclass[sigconf]{acmart}

\usepackage{tabularx}
\usepackage{multirow}

\AtBeginDocument{%
  \providecommand\BibTeX{{%
    \normalfont B\kern-0.5em{\scshape i\kern-0.25em b}\kern-0.8em\TeX}}}

\setcopyright{acmlicensed}
\copyrightyear{2018}
\acmYear{2018}
\acmDOI{XXXXXXX.XXXXXXX}

\acmConference[PrePrint for Conference acronym 'XX]{Make sure to enter the correct
  conference title from your rights confirmation email}{June 03--05,
  2018}{Woodstock, NY}

\acmISBN{978-1-4503-XXXX-X/18/06}

\begin{document}

\title{SketchConcept: Sketching-based Concept Recomposition for Product Design using Generative AI}

\author{Runlin Duan}
\orcid{0000-0001-8256-6419}
\authornote{Both authors contributed equally to this research.}
\affiliation{%
  \institution{School of Mechanical Engineering \\ Purdue University}
  \city{West Lafayette}
  \country{USA}}
\email{duan92@purdue.edu}

\author{Chenfei Zhu}
\orcid{0009-0003-3408-2876}
\authornotemark[1]
\affiliation{%
  \institution{School of Mechanical Engineering \\ Purdue University}
  \city{West Lafayette}
  \country{USA}}
\email{zhu1237@purdue.edu}

\author{Yuzhao Chen}
\orcid{0009-0005-6196-1176}
\affiliation{%
  \institution{Elmore Family School of Electrical and Computer Engineering \\ Purdue University}
  \streetaddress{610 Purdue Mall}
  \city{West Lafayette}
  \state{IN}
  \country{USA}
  \postcode{47907}
}
\email{chen4863@purdue.edu}

\author{Dizhi Ma}
\affiliation{%
  \institution{Elmore Family School of Electrical and Computer Engineering \\ Purdue University}
  \city{West Lafayette}
  \country{USA}}
\email{ma742@purdue.edu}

\author{Jingyu Shi}
\orcid{0000000151592235}
\affiliation{%
  \institution{Elmore Family School of Electrical and Computer Engineering \\ Purdue University}
  \city{West Lafayette}
  \country{USA}}
\email{shi537@purdue.edu}

\author{Ziyi Liu}
\affiliation{%
  \institution{School of Mechanical Engineering \\ Purdue University}
  \city{West Lafayette}
  \country{USA}}
\email{liu1362@purdue.edu}

\author{Karthik Ramani}
\orcid{0000-0001-8639-5135}
\affiliation{%
  \institution{School of Mechanical Engineering \\ Purdue University}
  \city{West Lafayette}
  \country{USA}}
\email{ramani@purdue.edu}


\begin{abstract}
Conceptual product design requires designers to explore the design space of visual and functional concepts simultaneously. 
Sketching has long been adopted to empower concept exploration. 
However, current sketch-based design tools mostly emphasize visual design using emerging techniques. 
We present SketchConcept, a design support tool that decomposes design concepts into visual representations and functionality of concepts using sketches and textual descriptions.
We propose a function-to-visual mapping workflow that maps the function descriptions generated by a Large Language Model to a component of the concept produced by image Generative Artificial Intelligence(GenAI).
The function-to-visual mapping allows our system to leverage multimodal GenAI to decompose, generate, and edit the design concept to satisfy the overall function and behavior.
We present multiple use cases enabled by SketchConcept to validate the workflow.
Finally, we evaluated the efficacy and usability of our system with a two-session user study.

\end{abstract}


\begin{teaserfigure}
  \includegraphics[width=\textwidth]{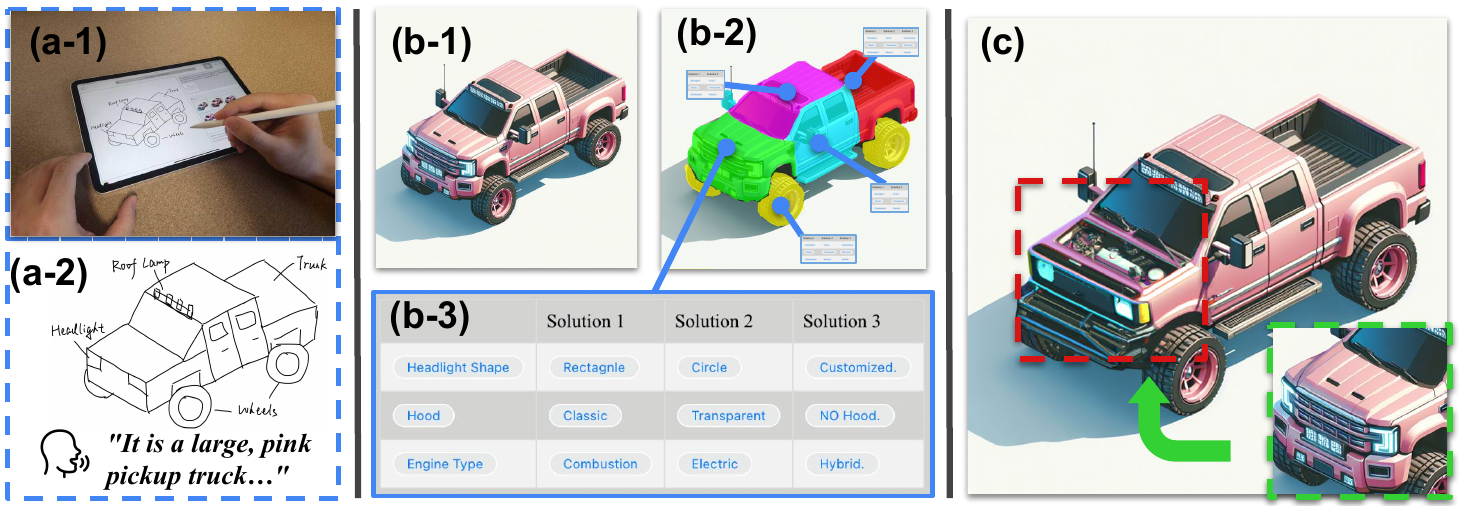}
  \caption{An overview of the SketchConcept system workflow. The user of SketchConcept wants to design a pink pickup truck. To generate a concept image using the SketchConcept system, the user first sketches the idea on the interface and describes the design intent naturally using voice input (a-1) and (a-2). The system analyzes user's sketch and voice inputs to produce a refined description and generates concept images (b-1). Then the system visually segments the generated image, analyzes, and maps functions of the design to different visual parts (b-2) and (b-3). User can now perform a component-level modification by choosing a new implementation of any function.}
  \Description{Enjoying the baseball game from the third-base
  seats. Ichiro Suzuki preparing to bat.}
  \label{fig:teaser}
\end{teaserfigure}


\maketitle

\section{Introduction}

Conceptual design represents the initial phase of the design process, where designers explore numerous concepts before converging to optimal solutions. 
To facilitate this process, concepts are often decomposed into manageable components for easier exploration and modification \cite{baxter2018product,piya2017co}. Traditionally, sketches and textual descriptions have been widely employed to empower such decomposition  \cite{ullman1990importance,song2004insights}. Built upon such representations, many prior works developed design support tools for conceptual design \cite{ghosh2019interactive,ILoveSketch,arora2018symbiosissketch,drey2020vrsketchin, han2022semantic,sarica2023design}. 
However, prior works consider the two representations separately, neglecting the synergistic potential of incorporating both of them.

The emergence of Large Language Model (LLM) and visual Generative AI (GenAI) significantly enhanced conceptual design by enabling language-based concept generation \cite{ma2023conceptual,di2022idea,zhu2023generative} and flexible image generation \cite{sketch2prototype,CollaborativeDiffusion,liu2022design,ko2023large,wu2023visual}. 
LLM can generate descriptions for overall concept or components to identify functional elements \cite{zhu2023biologically}. However, textual descriptions alone often fall short in conveying crucial visual details such as spatial arrangements, dimensions, and component relationships, making them less efficient for design workflows. 
On the other hand, visual GenAI has been adopted to generate and refine sketches from text prompts or freehand inputs, enhancing the visual aspect of conceptual design. 
While these systems exhibit promising visual representations, they often require extensive human effort on prompt engineering to accurately locate and modify the components of concepts \cite{wu2023visual}. This challege is often due to the failure of understanding the functional meanings of visual components, which are typically conveyed through textual descriptions. 
This limitation makes it challenging to modify the functional and visual aspects of a concept without inadvertently affecting the other.

To bridge the gap between functional and visual representations to support conceptual design, we introduce SketchConcept, an end-to-end multi-modal system that enables component-level concept generation and modification through concept decomposition, which breaks down the overall concept into components while effectively integrates functional and visual aspects. Through this process, the system provides a foundation for seamless component-level exploration and iteration. SketchConcept allows users to iteratively generate and refine components via a multi-modal interface that incorporates sketching, voice, and text inputs. We validated the system's effectiveness and usability through a preliminary study and a user study, demonstrating its potential to support iterative concept refinement.

Our contributions include:
\begin{itemize}
    \item SketchConcept, an end-to-end multi-modal design assistive system to support GenAI-based concept design with sketch and language.


    \item A preliminary and user study that validate the system's effectiveness and usability in supporting iterative conceptual design workflows.

\end{itemize}

\section{Related Work}

\subsection{Sketch Tools for Design and Sketch-based Interactions} \label{sec: SketchSupportTools}

With the advancement of touchscreen devices and paperless workflows, designers increasingly use sketch-based tools to express ideas~\cite{ILoveSketch, EverybodyLovesSketch, ShadowDraw, DrawFromDrawings, SymbiosisSketch}. These tools enhance the design process by supporting intuitive sketching and interaction. AI-powered tools further expand these capabilities by offering automatic sketch completion~\cite{ghosh2019interactive} and style-based transformations~\cite{EmoG, Cao_Yan_Shi_Chen_2019}.
Several works focus on sketch-driven UI creation~\cite{SketchingInterfaces, Mohian_2022} and symbol recognition for technical applications~\cite{HierarchicalParsing}. Additionally, sketch-based interactions allow intuitive design operations, such as drag-and-drop gestures for visualization~\cite{DataInk, ElicitingSketched, Suzuki_2020, SketchedReality, SketchStory}. Sketch-based retrieval systems have also been introduced, such as Xie et al.'s assembly component retrieval~\cite{Xie_2013} and Forte's 3D model generation~\cite{Forte}.
However, traditional sketch-based interactions heavily rely on users’ drawing skills, limiting their broader adoption. Our approach integrates generative AI (GenAI) to enhance sketch interactions, enabling users to generate high-quality designs without requiring expert sketching abilities.

\subsection{Conceptual Design with Generative AI}

Conceptual design involves iterative exploration and refinement of abstract ideas. GenAI aligns with this iterative process by leveraging existing data to generate novel concepts and facilitate creative exploration. Existing GenAI-driven conceptual design methods can be categorized into three approaches: \textit{text-based}, \textit{sketch-based}, and \textit{sketch with annotation}. 
\textbf{Text-based approaches} utilize knowledge graphs and text-mining techniques to support ideation and information retrieval~\cite{SemanticNetworks, Sarica_2020, DataDriven}. Visualization tools~\cite{DesignOpportunity} further assist in identifying design opportunities from text-based data. 
\textbf{Sketch-based methods} use models such as GANs~\cite{goodfellow2014generative} and VAEs~\cite{kingma2022autoencoding} to generate design alternatives from sketches~\cite{SynthesizingDesigns, ConditionalGenerative, chen2021beziergan, zhang20193d}. Additionally, CNN-based approaches enable 3D model retrieval from 2D sketches~\cite{wang2015sketchbased}.
\textbf{Sketch with annotation} integrates multimodal inputs, combining sketches with text to generate complex designs. Attention-based models~\cite{vaswani2017attention} power systems such as DreamSketch~\cite{DreamSketch} and Collaborative Diffusion~\cite{CollaborativeDiffusion}, enhancing creative flexibility.

In contrast to existing methods, SketchConcept enables component-level design refinement by integrating functional and aesthetic considerations, allowing designers to modify concepts through both sketching and textual inputs, streamlining the conceptual design workflow.

\begin{figure*}[h]
   \centering
   \includegraphics[width=.7\linewidth]{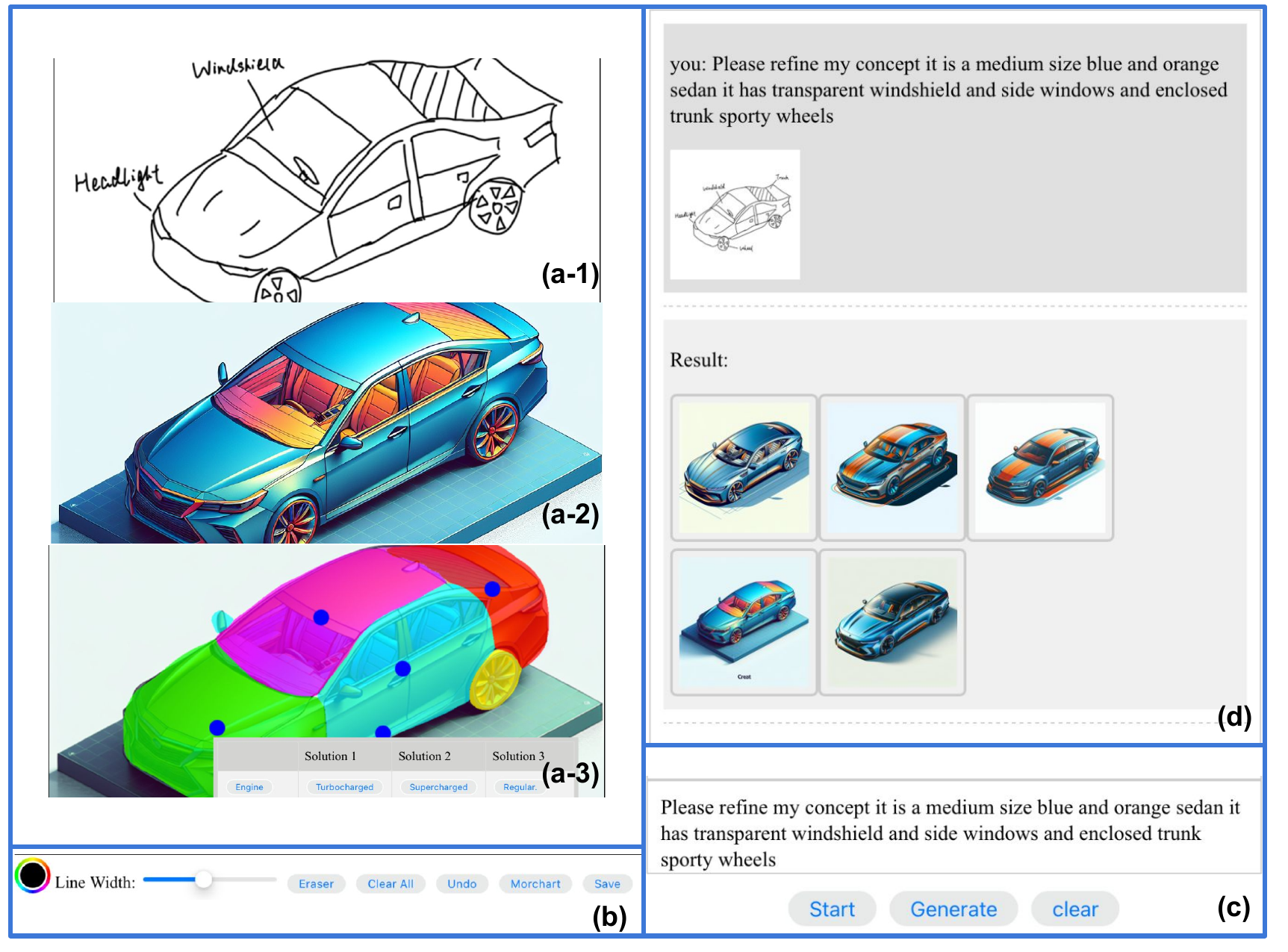}
   \caption{Interface layout of the SketchConcept. The user can draw (a-1), view (a-2), and edit (a-3) the concept image on the white canvas. The interface provides tools (b) to help the user sketch. The interface supports voice input (c), and the conversation between the user and the system will be recorded in the chat dialog (d). }
 \label{fig:ui}
\end{figure*}

\section{SketchConcept System} \label{sec:system}


In this section, we first introduce the design goal of SketchConcept. Then we provide an overview of SketchConcept system and introduce each module in detail.

\subsection{SketchConcept Design Goal}

SketchConcept is a GenAI-based assistive tool designed to support concept design by enhancing exploration and modification through the integration of functional and visual aspects at the component level. To achieve this, we define the following design goals for the SketchConcept system:

\begin{itemize}
    \item \textbf{Concept Generation (DG1).} Enable users to explore diverse concepts with GenAI supports by expressing their preferences through visual and textual inputs, fostering creativity and idea expansion.

    \item \textbf{Concept Decomposition and Visual-Function Alignment (DG2).} Decompose the overall concept into component-level elements while ensuring alignment and maintaining connections between visual and functional aspects.

    \item \textbf{Concept Modification (DG3).} Provide an intuitive way to refine concepts by remixing existing alternatives or generating new ones through visual and textual inputs, ensuring seamless iteration and adaptation.
\end{itemize}

\subsection{System Overview}
We developed the SketchConcept system to address the proposed design goals. SketchConcept is implemented as a sketching interface running on an iPad browser, providing an intuitive and interactive user experience. The system workflow consists of three key steps:

\textbf{Concept Generation:}  
The user sketches with a pen and provides voice input to convey their concept. The system analyzes the user's design intent and generates a set of concept images accordingly.

\textbf{Concept Decomposition:}  
The system semantically segments the concept image into components and maps functional elements to their corresponding visual representations. The functions of each component are organized into a function chart, where three alternative solutions are provided to support further exploration.

\textbf{Component-level Editing:}  
The system presents the segmented image alongside the function chart of each component. Users can edit their designs by redrawing a component or interacting with its function chart. The changes on one aspect will reflect on the other.

To illustrate the workflow, we present an example in which the user designs a pink pickup truck, as shown in ~\autoref{fig:teaser}.

\subsection{Concept Generation}
\label{subsec:conceptrefine}
The Concept Generation module enables users to explore various concepts by generating concept images based on their sketches and textual descriptions. The process begins with the user sketching their concept on the system interface and providing a voice description. The system analyzes the sketch and refines the design description using GPT-4V\footnote{https://openai.com/research/gpt-4v-system-card}, incorporating style and placement requirements to ensure consistency. The refined description is then used to generate a set of concept images via DALL-E3\footnote{https://openai.com/dall-e-3}, offering visual variations for exploration. 

The user selects their preferred concept image from the generated options. Next, the system analyzes the selected image and description to identify design functions and their corresponding solutions in the form of function-solution pairs (e.g., "wheel size - 19 inches" or "sunroof - panoramic").



\subsection{Concept Decomposition}
\label{subsec:conceptdecomp}
The Concept Decomposition module decomposes the GenAI-generated concept into component-level while ensuring alignment between visual and functional aspects. Concept decomposition consists of two key processes: visual decomposition and functional decomposition. 

For visual decomposition, a semantic segmentation neural network is employed to analyze concept images and identify distinct visual components. This is achieved by collecting and labeling a dataset of concept images, which is then used to train the segmentation model, as detailed in ~\autoref{sec:training}.

Given a concept image, the system applies the trained segmentation model to visually identify and divide components. It then maps the functions identified during concept generation to their corresponding visual components. Specifically, the system overlays each component with a distinct transparent color and leverages GPT-4V to determine the most likely color region associated with each function. Once the mapping is complete, a function chart is generated for each component, providing two additional alternative solutions to facilitate further exploration and modification.

\begin{figure}[h]
  \centering
  \includegraphics[width=1\linewidth]{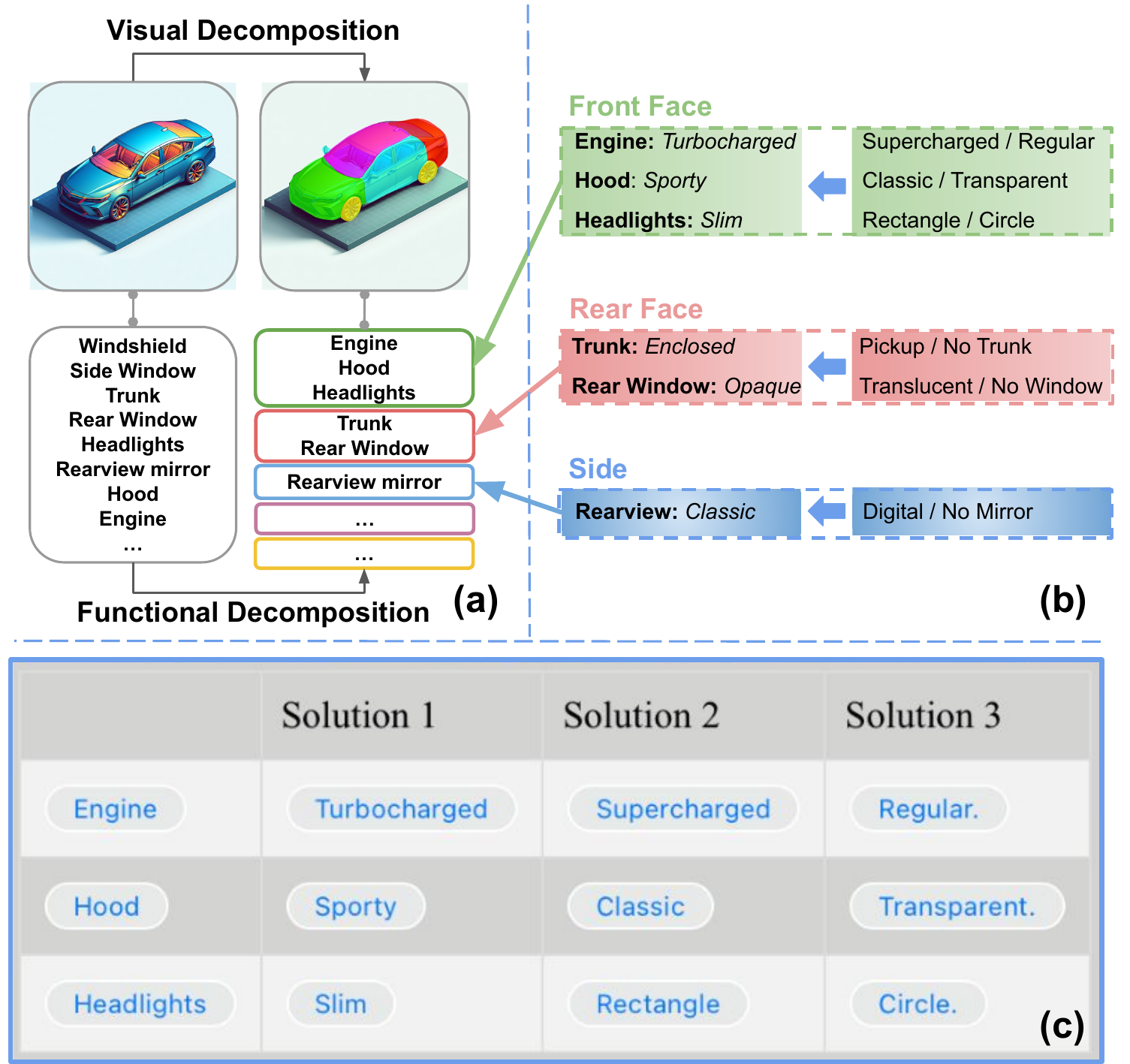}
  \caption{An example of concept decomposition. System segments the concept image semantically and maps the functions to components (a). Function charts are built by generating additional solutions for functions (b). The visualization of a function chart on the interface (c).}
  \label{fig:Concept Decomposition}
\end{figure}

An example of the concept decomposition process is illustrated in ~\autoref{fig:Concept Decomposition}. The interactive function chart assists users in exploring their designs by providing various functional alternatives.

\subsection{Component-Level Editing}

This module allows users to edit their concepts at the component level through multiple approaches, as illustrated in ~\autoref{fig:componentedit}.

\subsubsection{Editing by Recommendations}
As described in ~\autoref{subsec:conceptdecomp}, the system generates function charts during concept decomposition, each offering two alternative solutions per function. Users can modify a component by selecting an alternative solution from the chart. Upon selection, the system records the choice and utilizes GPT-4V to assess whether the change affects the component's visual representation using the prompt, "\textit{analyze if this function is visible in the image}". 
For instance, altering the shape of a car's headlights impacts its visual representation, whereas changing the engine material may not. If a visual change is detected, the system updates the corresponding component using DALLE-2 with the prompt, "\textit{change function from [SOLUTION\_A] to [SOLUTION\_B]}", such as "change wheel size from 19 inches to 20 inches".

\subsubsection{Editing by Sketching}
If the available alternatives do not fulfill the user’s design intent, they can manually edit the component through sketching. Users can open the function chart and click on the function to be modified instead of selecting a predefined solution. The system then displays the component area masked by a white canvas, enabling the user to sketch a new design and provide a voice description of their design intent. 

Using the same refinement method as in ~\autoref{subsec:conceptrefine}, the system processes the sketch and description to generate a refined prompt, which is then used to update the concept image via DALLE-2.


\begin{figure*}
  \centering
  \includegraphics[width=\linewidth]{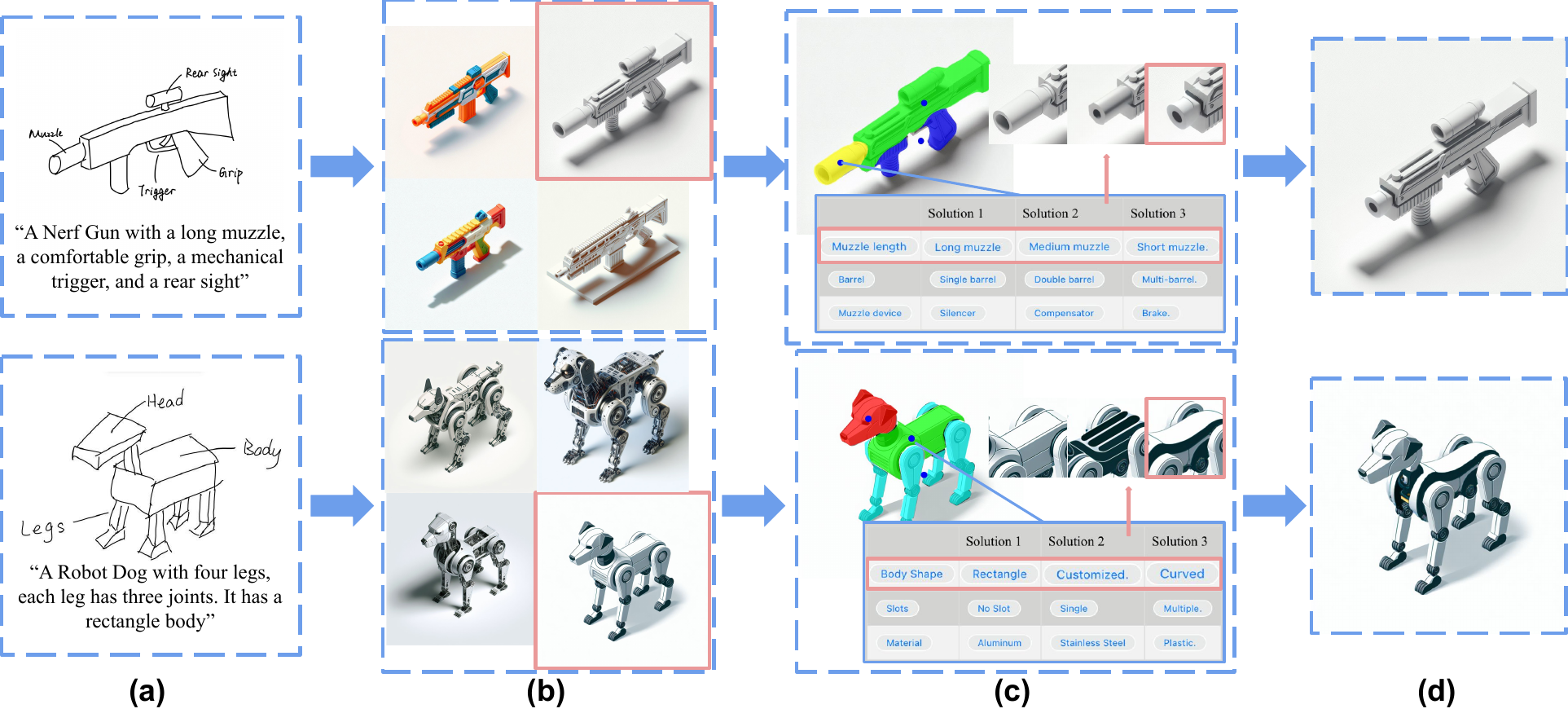}
  \caption{Conceptual toy designs of Nerf Gun and Robot Dog using SketchConcept system. (a) input sketch and description; (b) refined concept images; (c) fine-tune the concept with system-recommended solutions; (d) new concept images.}
  \label{fig:usecase1}
\end{figure*}

\section{Application Scenarios}

One application of the SketchConcept system is in toy design, which requires a combination of creativity and practical considerations related to mechanical and electronic sub-systems. Many toys are inspired by everyday objects and interactions, making them closely related to human experiences and requiring innovation to ensure they are engaging and appealing. Toy design applies product design principles across flexible scales and is widely practiced by individuals and industry professionals alike. Designers can create simple toys with intuitive mechanical interactions, such as windmills with rotational mechanisms, or incorporate advanced technologies to develop interactive virtual toys \cite{zhu2022mecharspace}.

To demonstrate the capabilities of SketchConcept, we present two toy design examples: a Nerf Gun and a Robot Dog. The Nerf Gun design focuses on mechanical interactions between sub-systems, while the Robot Dog design emphasizes the integration of electronic components. These examples showcase how SketchConcept assists designers by providing concept decomposition and component-level GenAI suggestions. As shown in \autoref{fig:usecase1}, the user begins by sketching an initial toy design (\autoref{fig:usecase1}-a), which the system generates into a group of concept images (\autoref{fig:usecase1}-b). The user can then modify the concept using system-recommended solutions (\autoref{fig:usecase1}-c, \autoref{fig:usecase1}-d). Currently, SketchConcept supports component and sub-system exploration through first-level and second-level concept decomposition, while lower-level hardware and software details are beyond the scope of conceptual design.

\begin{figure*}[h]
  \centering
  \includegraphics[width=\linewidth]{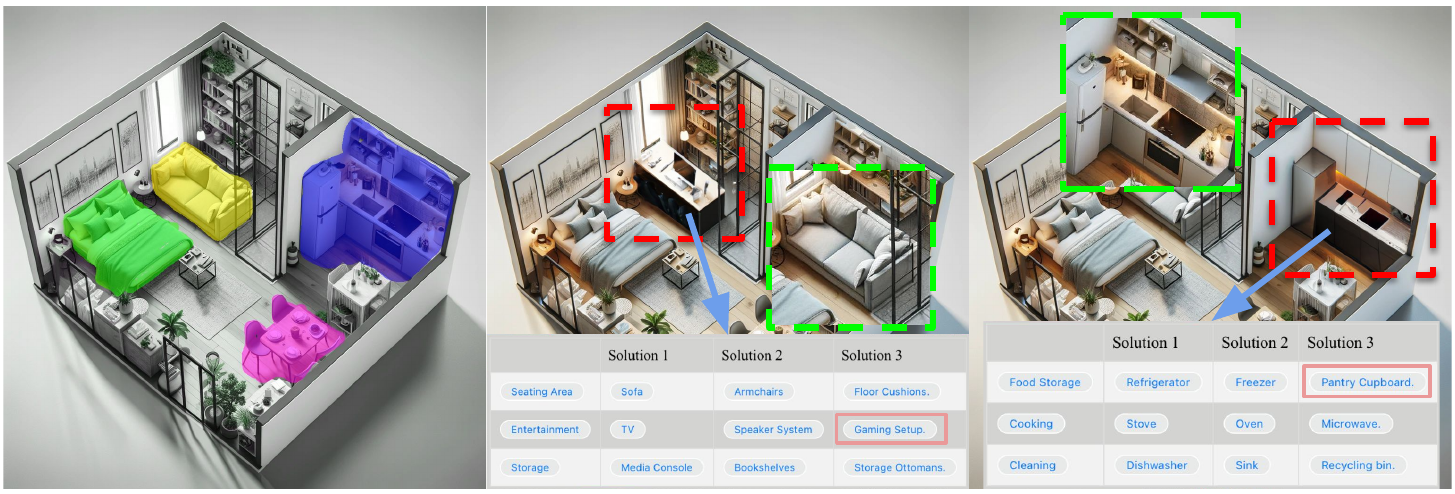}
  \caption{An example of editing a concept image of home layout in component level.}
  \label{fig:usecase3}
\end{figure*}

Beyond toy design, the SketchConcept system can be extended to other conceptual design domains, such as interior layout design. For instance, a tenant planning to renovate an apartment can use SketchConcept to create an initial room layout concept. The process starts with GenAI generating a layout, followed by system-assisted semantic decomposition (\autoref{fig:usecase3}-a). The system then provides component-level suggestions (\autoref{fig:usecase3}-b, \autoref{fig:usecase3}-c). By leveraging existing semantic segmentation datasets for real-world environments, SketchConcept can easily adapt to similar tasks by incorporating pre-trained segmentation models and optimizing GenAI performance through a human-GenAI collaboration framework.

These use cases demonstrate how SketchConcept facilitates conceptual design across various domains by enabling users to efficiently explore and refine component-level designs.

\section{Preliminary System for Evaluating Concept Decomposition}
\label{sec:training}

We collected a concept image dataset by annotating generated images from DALLE-3~\footnote{https://openai.com/dall-e-3} and implemented a semantic segmentation method to empower GenAI with the ability to capture visual representations of components.
We also conduct another preliminary study to test LLM's ability to understand the visual presentations of components.
In this section, we introduce and evaluate our method.

\subsection{Data Collection}
\label{sec:datacollection}
We chose three common design objects: a car, a Nerf Gun, and a Robot Dog.
We used DALLE-3 to generate concept images for each object.
To align the format of the generated contents, we prompted DALLE-3 to generate with "\textit{a realistic [Object\_Name] shown in an isometric perspective and in a clean background}".
Then, we annotated the images collected considering the functionality of the object.
We segmented the car into five components, Nerf Gun and Robot Dog into three components. We collected 600 images in total. 


\subsection{Semantic Segmentation}\label{sec:sem seg}
We split the whole dataset into a train set, a valuation set, and a test set with a ratio of 8:1:1, and also implement the data augmentation using random flipping and crop.

To perform semantic segmentation on our dataset, we selected 6 commonly used AI models. For each model, we use ResNet34~\cite{he2015deep} as the backbone, also choose multi-class dice loss as the loss function, Adam as the optimizer, and 1e-4 as the initial learning rate. At the same time, each model uses the pre-trained weight trained on the ImageNet dataset~\cite{Imagenet}. Since the amount of data for each category is relatively small, we train those models with 40 epochs and perform a learning rate decay of 0.1 at the 25th epoch.

Each model has its own architecture and is suitable for different scenarios. Therefore, we measure the performance of each model using mean IoU:

\begin{equation}
    \text{mean IoU} = \frac{1}{N} \sum_{i=1}^{N} \frac{\text{TP}_i}{\text{TP}_i + \text{FP}_i + \text{FN}_i}
\end{equation}

where N denotes the total number of classes for each category. TP, FP, and FN denote true positive, false positive, and false negative rates respectively. Intuitively, this metric represents the overlap ratios between the predicted area and the ground truth area. It ranges from 0 to 1, higher value means better performance. Results list in Table\ref{tab:seg_per}

\begin{table}[ht]

\begin{center}
\footnotesize
\captionof{table}{mean IOU for each models on our datasets} \label{tab:seg_per}
\begin{tabular}{lcccccc}
\hline
Methods & Car & NerfGun & RobotDog\\
\hline
FPN\cite{kirillov2019panoptic} & 0.940 & 0.928 & 0.916\\
UNet\cite{ronneberger2015unet} & 0.920 & {\bf0.938} & 0.935\\
UNet++\cite{zhou2018unet} & 0.934 &  {\bf0.938} & 0.934 \\
DeepLabV3\cite{chen2017rethinking} & 0.947 &{\bf0.938} & {\bf0.937} \\
DeepLabV3+\cite{chen2018encoderdecoder} & 0.941 & 0.921 & 0.896 \\
PAN\cite{li2018pyramid} & {\bf0.951} & 0.927 & 0.905 \\

\hline

\end{tabular}
\end{center}

\end{table}

\subsection{Function Mapping}
To evaluate ChatGPT's ability to understand the semantic meaning of different parts in concept design, we chose a set of functions and used ChatGPT to map the functions to the visual parts. Due to the variety of designs generated by DALLE-3 and potential users, we only test the functions that are essential and relatively fixed in the concept design. We will discuss those functions whose dependencies are more likely to be influenced by design in User Study. For each design class, 6-10 functions are selected. We then calculate the accuracy of GPT-4V in classifying these functions on our dataset to evaluate its performance in understanding the semantic meaning of concept images.
\begin{table*}[h]
\centering
\footnotesize
\captionof{table}{Qualitative results of function mapping on three design objects in our dataset. Each object is tested with seven common functions. The number reported in the table represents the accuracy of GPT-4V in mapping functions to their visual counterparts.} \label{tab:mappingtable}
\begin{tabular}{ccccccccc}

\hline
 \textbf{Category} & \multicolumn{7}{c}{\textbf{Function}} \\
\hline
\multirow{2}{*}{Car} & headlight & front bumper  & windshield & mirror & trunk & tire & suspension\\ 
    & 100       & 100             & 84.7        & 98.5    & 100   & 100  & 89.5        \\ \hline
\multirow{2}{*}{Nerf Gun} & muzzle & barrel & rear sight & safety & trigger & gripper & magazine \\  
        & 100    & 81.2   & 97.6       & 85.4   & 100     & 90.2    & 80.7          \\  \hline
\multirow{2}{*}{Robot Dog} & eyes & ears & load & tail & hip & ankle & claw   \\  
          & 100    & 100   & 100       & 100   & 87.5     & 100    & 100            \\ \hline
\end{tabular}
\end{table*}

\subsection{Results and Discussion}
The results are summarized in Tables 1 and 2. 
All the six models evaluated performed good on segmenting the concept images of Car, Nerf Gun, and Robot Dog by achieving the highest mean IOU of 0.951 and the lowest of 0.896, demonstrating the generalizability of the dataset and the task. Among the six models, DeepLabV3\cite{chen2017rethinking} achieved overall best performance by reaching the mean IOU of 0.947, 0.938, and 0.937 on Car, Nerf Gun and Robot Dog, respectively. Thus, we chose DeepLabV3 as our segmentation model in our system.

On the function mapping, GPT-4V shows strong abilities to identify the functional meaning of the components from their visual representations, achieving the highest mapping accuracy of 100\% and the lowest accuracy of 80.7\% among the three classes. Functions that are easy to identify and are more fixed in previous designs are more likely to be mapped correctly by GPT-4V.
We discuss other failure cases of the method in ~\autoref{sec:fw}. Overall the results show good performances on both semantic segmentation and function mapping.

\section{User Study}

We conducted a user study to evaluate overall system usability and the efficacy of Sketchconcept. 
Ten users were recruited (six males and four females, aged 19-23).
They are all mechanical engineering undergraduates who have been introduced to the principles of product design.
None of the users had experienced our system before conducting the user study. 
The entire user study took around one hour, and each participant was reimbursed 20 dollars for their time. 
After the study, the user completed a survey with Likert-scale questions (scaled 1-5) regarding the user experiences with specific system features and a standard System Usability Scale (SUS) questionnaire.
We conducted an open-ended interview to get subjective feedback on our system. 
The user study was under the approval of a university IRB. 

\begin{figure}[h]
  \centering
  \includegraphics[width=1\linewidth]{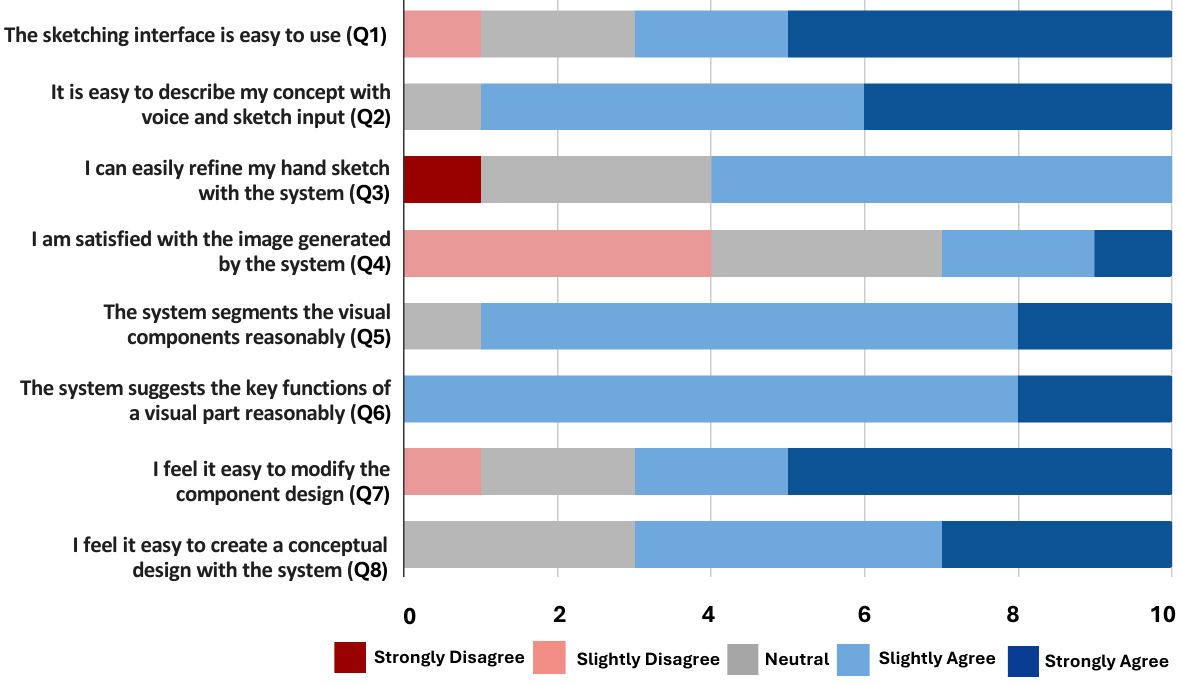}
  \caption{The results of the qualitative feedback on the system usability.}
  \label{fig:USER1}
\end{figure}

\subsection{\textit{Results and Discussion}} 
The Likert-type question ratings are shown in \autoref{fig:USER1}. 
In general, users found the system easy to use (Q8: avg=4, std=0.82), and the sketching interface is clear (Q1: avg=4.1, std=1.1). \textit{"I thought the sketching interface was very intuitive (P4)"} 
Describing concepts using the sketch-voice input provided by our system is also considered straightforward and simple (Q2: avg=4.4, std=0.70).
The main features of SketchConcept are also generally recognized by participants.
Most of them acknowledged that the concept decomposition feature performed well, providing reasonable image segmentation (Q5: avg=4.1,std=0.57) and function mapping (Q6: avg=4.2,std=0.42), and they could do component-level modifications on their concepts easily based on this decomposition (Q7: avg=4.1, std=1.1).\textit{ I like the segmentation and the morphological chart that helps me modify the concept. (P6)}

For the concept refinement feature, more than half of the participants think the system provides simple interaction for refining the concept (Q3: avg=3.4, std=0.97). \textit{There's a refinement process that allowed me to not make modifications on the entire image, I was really impressed by how it mostly kept. (P4)} One participant struggled with the feature because the participant encountered a copyright problem when prompting GenAI.
Not all participants agree that the system can generate satisfactory images (Q3: avg=3.0,std=1.1). 
One problem that was complained about the most was that when the design intent is very complex or it is difficult to find a prototype in reality, the system is not able to generate reasonable images. \textit{"It was a little more difficult with the Nerf Gun. 
The thought in my head was a Nerf Gun that I had at home. It is not that common, because it has a manual system. (P3)"}

\begin{figure*}[ht]
  \centering
  \includegraphics[width=\linewidth]{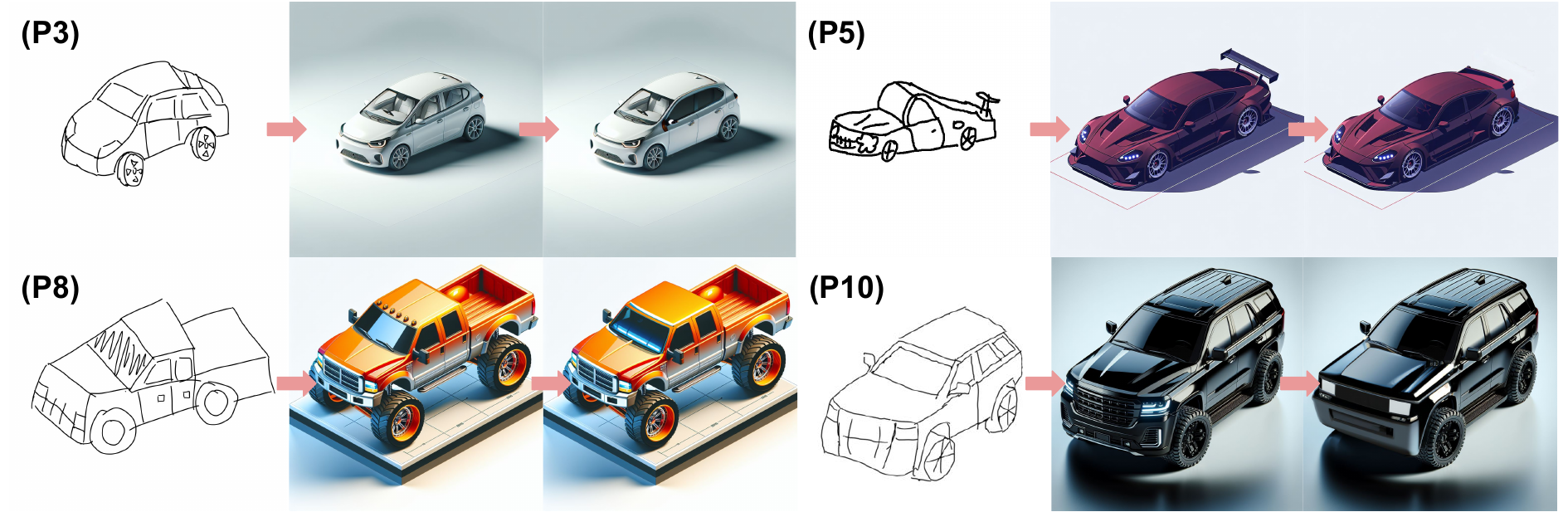}
  \caption{Car concept design outcomes from the second session of user study.}
  \label{fig:userwork}
\end{figure*}

We also showcase the concepts created by selected users during the user study, as shown in \autoref{fig:userwork}. Each user's concept image was generated with at least three design intents and was fine-tuned at least once.

\section{Limitation and Future Works}
\label{sec:fw}

\subsection{Exploring Design Components with GenAI}

The result of our user study shows that our system allows users to efficiently explore design components by adopting the suggestions provided by GenAI. 
\textit{"The best feature of this system is definitely segment-based editing rather than having to regenerate the entire image. (P9)"}
This is largely attributed to the semantic-to-visual mapping problem solved by the system.
By mapping the textual descriptions of the components to the visual representations, the system can suggest modification of components based on the functions description generated by LLM. Meanwhile, our system can use vision-based GenAI to generate the visual representation of the components, based on the changes in the function descriptions.
On the interface end, we provide the user with the function explorer to facilitate the semantic-to-visual mapping operation.
With the component explorer, the user not only gets the choices on different components but can also visualize the concepts after modification.
\textit{I like the component explorer, I can see how my design would look after I have made it and if I change different features. (P6)}
The use of visual and textual stimuli has long been considered an efficient way to enhance design creativity and efficiency \cite{luo2021guiding,han2020data}. 
The users also mentioned that exploring the concept component-wise with GenAI could help them think out-of-the-box. 
\textit{"It segments out each part into like individual things. There were things I didn't even think of that were suggested. (P4)"
}

\subsection{Create Dataset for Product Design}

The system achieves reasonable visual and functional decomposition on a concept by adopting a semantic segmentation model fine-tuned on our product dataset.
\textit{I think it does a very good job overall, segmenting everything of the concept in images and texts. (P3)}
While the user is satisfied with the segmentation result provided, they suggest that further segmentation in the lower-level component would be helpful in further detailed design.
\textit{"Getting the segmentation down to very small parts would be great to have. (P4)"
}
Decomposing a product into components with visual representation and functional descriptions makes it possible to describe, modify, and organize the product using the emerging LLM techniques.
Recent work has focused on creating task-specific data sets for engineering design purposes, such as exploring the engineering parts \cite{ramanujan2015framework,kim2020large} or processing the geometric elements \cite{seff2020sketchgraphs,seff2021vitruvion}.
To achieve detailed segmentation, a product design data set is required to collect a wide range of products with semantic information on functional meanings.

\subsection{Collaboration Design for Human-GenAI system}

All users can achieve their design goals with the assistance of GenAI. Nevertheless, we have observed that in certain situations, GenAI struggles to comprehend and implement the design intentions of humans.
\textit{"I tried to draw the Nerf Gun, it keeps changing my sketch to a firing cannon instead of a Nerf Gun. (P5) "
}
\textit{I draw a Nerf Gun with a magazine coming on the left and it gives me a Nerf Gun for the initial sketch that initially generated the image with the magazine down (P6).
}
To address this problem, it is necessary to introduce system functionalities that improve the interaction between human-GenAI, like refining prompts and correcting feedback, to guarantee the effectiveness of the system.

\section{Conclusion}
In this work, we present SketchConcept, a GenAI-based concept generation, decomposition, and editing tool that enables designers to refine and edit their design intents.
SketchConcept utilizes the ability of two GenAI-based modules to accomplish textual descriptions and visual representations of design concepts.
To enable component-level editing of concept design, we collected and annotated AI-generated concept design images of three common design objects, and implemented a semantic segmentation method to empower GenAI with the ability to understand the visual representation of design functions. Based on the visual segmentation and function mapping, we perform component-level editing on the concept image. We demonstrate three groups of application scenarios. Through a
two-session user study, we first proved our system’s usability and then its utility as a design tool to help generate and edit concept designs. 

\bibliographystyle{ACM-Reference-Format}
\bibliography{citations}

\appendix

\end{document}